\documentclass{article}

\usepackage[preprint]{neurips_2019}
\usepackage{amsmath}
\numberwithin{equation}{section}
\usepackage{CJKutf8}
\usepackage[T1]{fontenc}    
\usepackage{hyperref}       
\usepackage{url}            
\usepackage{booktabs}       
\usepackage{amsfonts}       
\usepackage{nicefrac}       
\usepackage{microtype}      
\usepackage{graphicx}
\usepackage{epstopdf}
\usepackage{setspace}
\usepackage{rotating}
\usepackage{subfig}
\usepackage{verbatim}
\usepackage{amssymb}
\usepackage{amsmath}
\usepackage{algorithm2e}
\usepackage{verbatim} 
\usepackage{epstopdf}
\usepackage{setspace}
\usepackage{amsthm}
\usepackage{rotating}
\usepackage[T1]{fontenc}
\usepackage{subfig}

\usepackage{comment}
\usepackage{hyperref}
\usepackage{tikz,xcolor,hyperref}
\usepackage{caption} 
\usepackage{hyperref}
\usepackage{setspace}
\usepackage{soul}
\usepackage{lineno}
\usepackage{listings}
\usepackage{color}
\usepackage{verbatim}
\definecolor{mygreen}{rgb}{0,0.6,0}
\definecolor{mygray}{rgb}{0.5,0.5,0.5}
\definecolor{mymauve}{rgb}{0.58,0,0.82}
\definecolor{lime}{HTML}{A6CE39}
\DeclareRobustCommand{\orcidicon}{
	\begin{tikzpicture}
	\draw[lime, fill=lime] (0,0) 
	circle [radius=0.16] 
	node[white] {{\fontfamily{qag}\selectfont \tiny ID}};
	\draw[white, fill=white] (-0.0625,0.095) 
	circle [radius=0.007];
	\end{tikzpicture}
	\hspace{-2mm}
}
\foreach \x in {A, ..., Z}{%
	\expandafter\xdef\csname orcid\x\endcsname{\noexpand\href{https://orcid.org/\csname orcidauthor\x\endcsname}{\noexpand\orcidicon}}
}

\usepackage[compact]{titlesec} 
\title{Evolutionary trend of SARS-CoV-2 inferred by the homopolymeric nucleotide repeats}
\author{%
	Changchuan Yin \orcidA \thanks{Correspondence author, cyin1@uic.edu} \\\\
	Department of Mathematics, Statistics, and Computer Science \\
	University of Illinois at Chicago \\
	Chicago, IL 60607 \\
	USA \\
}
\begin{document}
\maketitle
\begin{abstract}
Severe acute respiratory syndrome coronavirus 2 (SARS-CoV-2) is the causative agent of the current global COVID-19 pandemic, in which millions of lives have been lost. Understanding the zoonotic evolution of the coronavirus may provide insights for developing effective vaccines, monitoring the transmission trends, and preventing new zoonotic infections. Homopolymeric nucleotide repeats (HP), the most simple tandem repeats, are a ubiquitous feature of eukaryotic genomes. Yet the HP distributions and roles in coronavirus genome evolution are poorly investigated. In this study, we characterize the HP distributions and trends in the genomes of bat and human coronaviruses and SARS-CoV-2 variants. The results show that the SARS-CoV-2 genome is abundant in HPs, and has augmented HP contents during evolution. Especially, the disparity of HP poly-(A/T) and ploy-(C/G) of coronaviruses increases during the evolution in human hosts. The disparity of HP poly-(A/T) and ploy-(C/G) is correlated to host adaptation and the virulence level of the coronaviruses. Therefore, we propose that the HP disparity can be a quantitative measure for the zoonotic evolution levels of coronaviruses. Peculiarly, the HP disparity measure infers that SARS-CoV-2 Omicron variants have a high disparity of HP poly-(A/T) and ploy-(C/G), suggesting a high adaption to the human hosts.
\end{abstract}
\textbf{{\large keywords}}: SARS-CoV-2, 2019-nCoV, homopolymeric nucleotides, zoonotic evolution, genome

\section*{Highlights}
\begin{itemize}
	\item SARS-CoV-2 variants contain an overabundance of C>T mutations.
	\item The C>T mutations in SARS-CoV-2 variants imprint the SARS-CoV-2 variant genomes by APOBEC cytosine deaminases in immune response.
	\item The disparity of homopolymeric nucleotide repeats poly-(A/T) and ploy-(C/G) is the result of C>T mutations.
	\item The disparity of homopolymeric nucleotide repeats poly-(A/T) and ploy-(C/G) may infer the trend of virus evolution.  
\end{itemize}

\section{Introduction}
\label{Introduction}
The novel coronavirus SARS-CoV-2, the causative agent for Coronavirus disease-2019 (COVID-19) pandemic, has infected 276.436 million individuals and caused 5.374 million deaths in the globe as of Dec. 23, 2021 \citep{who_2021}. Understanding the evolutionary origin and trend of SARS-CoV-2 is of importance for controlling the current COVID-19 pandemic, designing effective vaccines and antiviral drugs, and preventing the future epidemic of coronavirus diseases. 

To gain a comprehensive zoonotic evolution of SARS-CoV-2, we contextualize SARS-CoV-2 in bat and human coronaviruses. Coronaviruses are enveloped positive-strand RNA viruses with crown-like spikes on their surface. Coronaviruses belong to the family\textit{Coronaviridae} and the order \textit{Nidovirales}. Coronaviruses can widely spread in humans, other mammals, and birds. Human coronaviruses (HCoVs) were first identified in the mid-1960s. Seven common HCoVs are CoV-229E (alpha coronavirus), CoV-NL63 (alpha coronavirus), CoV-OC43 (beta coronavirus), CoV-HKU1 (beta coronavirus), Severe acute respiratory syndrome coronavirus (SARS-CoV), Middle East respiratory syndrome coronavirus (MERS-CoV), and current SARS-CoV-2. CoV-229E and CoV-OC43 are the cause of the common cold in adults during the mid-1960s. CoV-HKU1 and CoV-NL63 cause disease manifestations include the common cold and chronic pneumonia. CoV-HKU1 has been predominantly reported in children in the United States but less common among adults. Three highly pathogenic human coronaviruses, SARS-CoV, MERS-CoV, and SARS-CoV-2, which emerged in 2002, 2012, and 2019, respectively, have caused severe respiratory diseases \citep{chen2020pathogenicity}. The estimated fatality rate in the confirmed cases is 6.6\% in SARS-CoV-2, which is lower than those of SARS-CoV and MERS-CoV, 9.6\% and 34.3\%, respectively \citep{wang2020novel}.

Science consensus admits that SARS-CoV-2 coronavirus originated in bats, specifically horseshoe bats. A few closely related to SARS-CoV-2 have been identified. Bat coronavirus RATG13 was isolated from \textit{Rhinolophus affinis} (a horseshoe bat) in Yunnan China in 2013. RATG13 genome is 96\% identical to that of SARS-CoV-2 \cite{zhou2020pneumonia}. \textit{R. malayanus} BANAL-52 is the closest known relative of SARS-CoV-2 found in Laos \citep{temmam2021coronaviruses}. BANAL-52 bat coronaviruses exhibit a high nucleotide identity with SARS-CoV-2 throughout the whole genome. The next-closest coronavirus is RmYN02, found in Malayan horseshoe bats (\textit{Rhinolophus malayanus}), which shares 93\% of its genetic sequence with SARS-CoV-2 \citep{zhou2020novel}. RmYN02 possesses high sequence similarity to SARS-CoV-2 of known\textit{ Sarbecoviruses} for most of its genome. Malayan pangolins (\textit{Manis javanica}) share up to 92\% of their genomes with the coronavirus SARS-CoV-2. The studies suggest that pangolins can host coronaviruses that share a common ancestor with SARS-CoV-2, but do not confirm that pangolins are the immediate host of SARS-CoV-2. The originated source and intermediate host animals of SARS-CoV-2 remain unconfirmed. The intermediate hosts of the coronaviruses can be bats, rodents, rabbits, and hedgehogs \citep{monchatre2017identification}. Since SARS-CoV was founded in 2003, a few researches have identified closely related bat-CoVs, including SLCoV/ZXC21 and SLCoV/ZC45 \citep{hu2017discovery}. Despite numerous phylogenetic analyses on these closely related CoVs, the zoonotic origin and timing of SARS-CoV-2 are still not confirmed. Though similarities of the genomes of these CoVs provide some clues on the zoonotic evolution, the evolutionary orders of these CoV are not fully established.  

When a virus infects an animal or human host, the host immune system recognizes the foreign infection and strives to eliminate the virus. Immune monitoring and abolishing the viral infectivity infection can be achieved at both protein and nucleic acid levels. As the innate immune antiviral response at nucleic acid level, the interferon-stimulated APOBEC (apolipoprotein B mRNA editing enzyme, catalytic polypeptide-like) RNA editing system recognizes the virus genomes and catalyzes Cytosine deamination to Uracils (C>U) in the virus genomes, leading to inhibit viral infectivity \citep{bishop2004apobec}. Recent studies suggested that the high-frequency C>U mutation mediated by APOBECs RNA editing is important in the rapid evolution and adaptation of the SARS-CoV-2 coronavirus in hosts \citep{wang2020host,di2020evidence,matyavsek2020mutation}. The host adaptation of the virus infection results in the accumulation of homopolymeric repeats of Uracils (ploy-Us, equally, poly-Ts in sequence notation) in the virus genomes. Therefore, the homopolymeric repeat poly-Ts are the fighting hallmarks of the immune system on the virus genomes. 

\cite{denver2004abundance} pioneered the study on homopolymeric repeat distribution in \textit{Caenorhabditis elegans} genome. The study showed that the frequencies and length distributions of A and T homopolymers vastly outnumber G and C homopolymers. The homopolymeric repeats signal the evolution of genomes \cite{denver2004abundance}. The other extensive studies revealed that Simple Sequence Repeats (SSRs), including homopolymeric nucleotide repeats, have been associated with antigenic variation and other adaptation strategies in Gram-negative organisms \citep{janulczyk2010simple}.

In this study, from sequence analysis, we show that the SARS-CoV-2 genome is abundant in homopolymeric nucleotide repeats. We here explore the evolutionary implications and significance of these repeats in the zoonotic shift. We propose to employ the homopolymeric nucleotides as the genome signature for measuring the zoonotic trend and orders of coronaviruses. 

\section{Methods}
\subsection{Measurement of the disparity of homopolymeric nucleotide repeats}
The changes of genomic structures and compositions of a virus are the result of the evolutionary adaption of the virus to hosts. To estimate the evolutionary trends and orders of coronaviruses from the genomic HP composition, we propose a measure of the HP disparity or HP compositions as follows. Let $HP(\alpha _i),\alpha  \in \{ A,T,C,G\} ,i = 3,4,5$ denote the HP repeat frequencies of a genome, the HP disparity, $R_{HP(AT/CG)}$  is defined as the frequencies of $HP(A_{3,4,5})$ and $HP(T_{3,4,5})$ over the frequencies of $HP(C_{3,4,5})$ and $HP(G_{3,4,5})$ (Equation 2.1). The homopolymeric nucleotides examined are only from 3-mers to 5-mers. It is noted that the n-mers within the m-mer (m>n) are not included when counting in the frequency of n-mers in Equation 2.1. For example, in DNA sequence, $T_{5}$ contains 3 $T_{3}$ and 2 $T_{4}$, when counting frequencies in Equation 2.1, the frequency $T_{5}$ is 1, and the frequencies of $T_{3}$ and $T_{4}$ are zero.
\begin{equation}
R_{HP(AT/CG)}  = \frac{{f(HP(A_{3,4,5} ) + HP(T_{3,4,5} ))}}
{{f(HP(C_{3,4,5} ) + HP(G_{3,4,5} ))}}
\end{equation}
, where $f( \cdot )$ is the frequency of a homopolymeric nucleotide repeat.

Another genomic composition measure is the ratio of A+T and C+G compositions, which is defined as A+T composition over C+G compositions (Equation 2.2).
\begin{equation}
R_{(AT/CG)}  = \frac{{f(A) + f(T)}}{{f(C) + f(G)}}
\end{equation}

\subsection{Genome data}
The complete genomes of coronaviruses, including SARS-CoV-2 \citep{wu2020new}, Severe Acute Respiratory Syndrome coronavirus (SARS-CoV), SARS-like coronaviruses (SL-CoVs), Middle East Respiratory Syndrome coronavirus (MERS-CoV), coronaviruses found in bats (bat-CoVs) or pangolins (Pangolin-SLCoV), and human coronaviruses (Human-CoVs), were retrieved from NCBI GenBank or GISAID database \citep{shu2017gisaid}. The complete genomes of SARS-CoV-2 isolates from the infected individuals were retrieved from the GISAID database between Jan.1, 2020 and Dec.4, 2021. Only the complete genome sequences without uncertain nucleotide 'N' of high-coverage were included in the test datasets. The NCBI GenBank and GISAID access numbers of the complete genomes are provided in the result tables and supplementary materials. 

\section{Results}
\subsection{SARS-CoV-2 genome has abundant HPs}
We assessed the HP distributions and contents in the genomes of SARS-CoV-2, SARS-CoV coronaviruses. The HP distribution results demonstrate that the SARS-CoV genomes have an abundance of HPs (Fig.1), compared with human endogenous retroviruses (HERV) \citep{nelson2003demystified} (Fig.2). The reason for using HERV as HP reference is that human endogenous retroviruses have been proposed to establish a protective effect against exogenous viral infections \citep{nelson2003demystified}. The polynucleotide mer length in SARS-CoV-2 genome can reach $poly-{T_5}$ and $poly-{G_4}$. Especially, in CoV genomes, the contents of poly-and poly-A are much richer than ploy-C and poly-G. but this disparity of poly-(A/T) and poly-(C/G) is not found or insignificant in other virus genomes, for example, HERV (Fig. 2), human immunodeficiency virus (HIV) and the hepatitis C virus (HCV), and Ebola (data not shown). In addition, SARS-CoV-2 genome has increased poly-(A/T) tracts, compared with SARS-CoV. According to these observations, we propose the definition of HP disparity score (Equation 2.1) to measure the biased content of poly-(A/T) vs poly-(C/G).

\begin{figure}[tbp]
	\centering
	\subfloat[]{\includegraphics[width=2.5in]{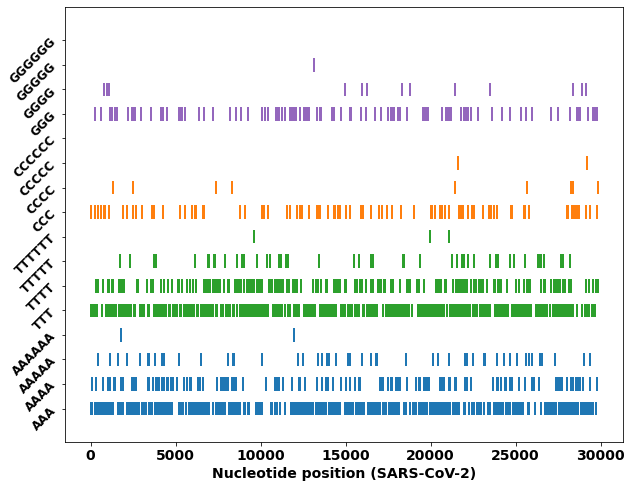}}
	\subfloat[]{\includegraphics[width=2.5in]{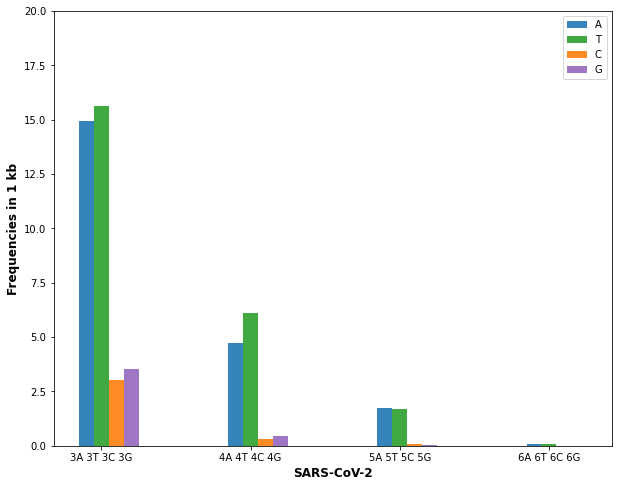}}\quad
    \subfloat[]{\includegraphics[width=2.5in]{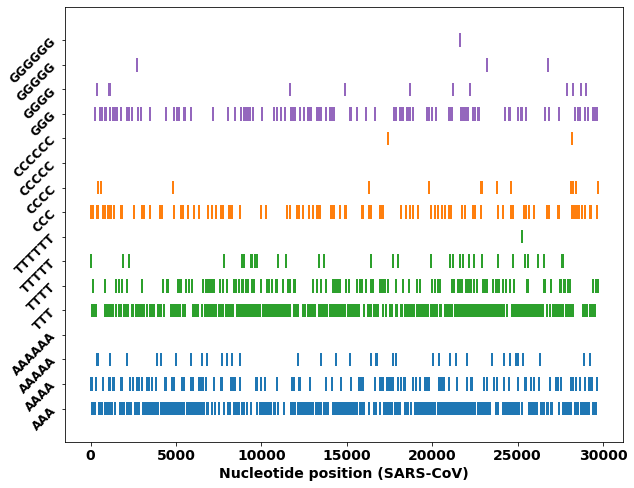}}
    \subfloat[]{\includegraphics[width=2.5in]{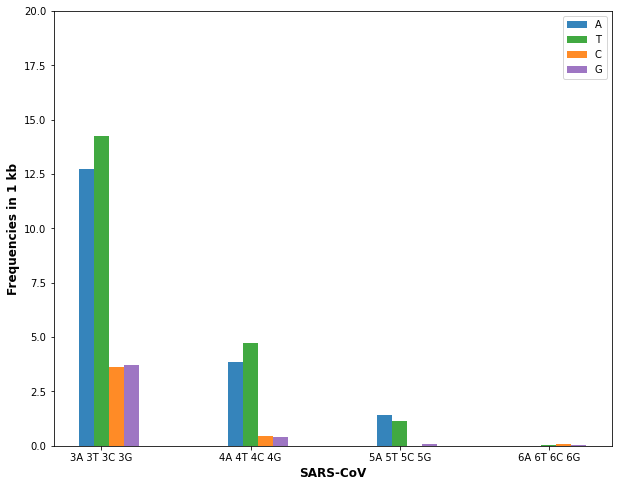}}\quad
	\caption{Distributions of HPs in SARS-CoV genomes. (a) HP distribution in SARS-CoV-2 genome (GenBank: NC045512), (b) HP frequencies in SARS-CoV-2 genome,  (c) HP distribution in SARS-CoV genome (GenBank: AY274119), (d) HP frequencies in SARS-CoV genome.}
	\label{fig:sub1}
\end{figure}
\begin{figure}[tbp]
	\centering
	\subfloat[]{\includegraphics[width=2.5in]{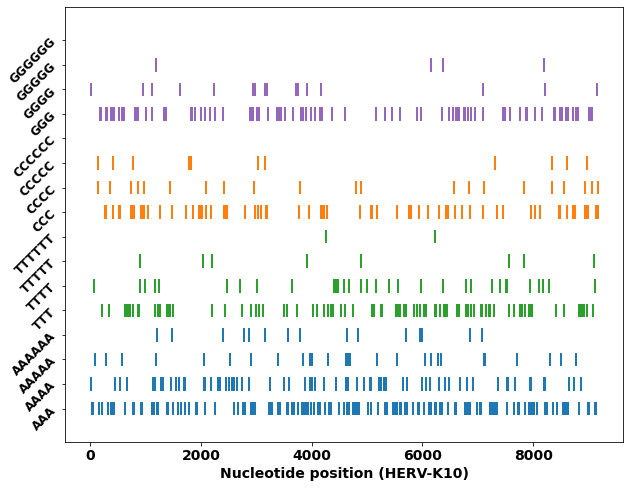}}
	\subfloat[]{\includegraphics[width=2.5in]{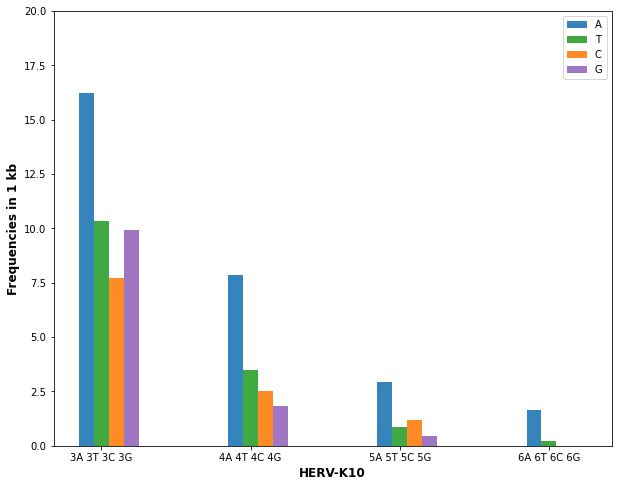}}\quad
	\subfloat[]{\includegraphics[width=2.5in]{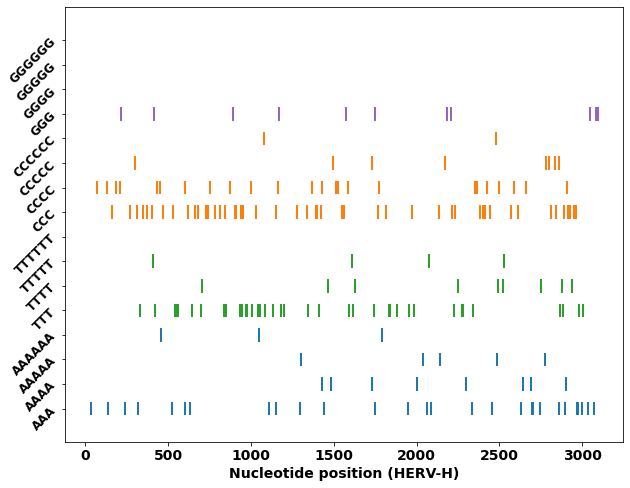}}
	\subfloat[]{\includegraphics[width=2.5in]{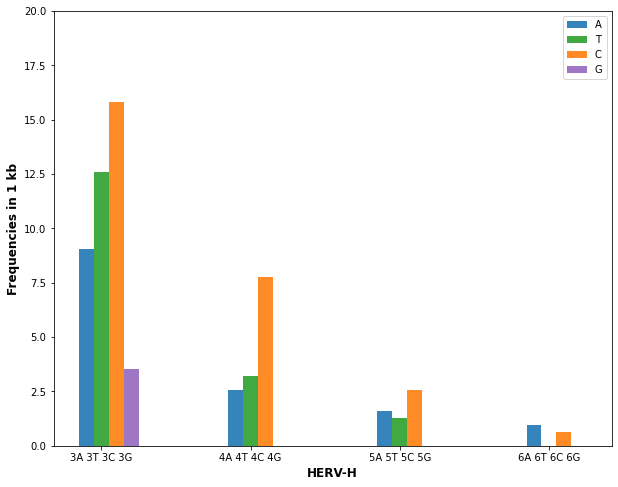}}\quad
	\caption{Distributions of HPs in HERV genomes. (a) HP distribution in HERV-K10 genome, (b) HP frequencies in HERV-K10 genome (GenBank: M14123),  (c) HP distribution in HERV-H genome (GenBank: AH007766), (d) HP frequencies in HERV-H genome.}
	\label{fig:sub1}
\end{figure}

\subsection{HPs in a CoV genome are the imprint of zoonotic evolution}
To investigate the correlation of HPs in the coronavirus genomes and the zoonotic evolution of coronaviruses, we measure the HP repeat disparity scores $R_{HP(AT/CG)}$ and AT/CG composition ratios $R_{(AT/CG)}$ (Equations 2.1 and 2.2) of the coronavirus genomes (Table 1). These two measures estimate the abundance and distribution bias of homopolymeric nucleotides. We also review the association of the HP disparity scores with zoonotic and pathogenicity characteristics.

The HP computation result shows that if a coronavirus can infect the human host with high pathogenicity, its HP disparity score is low, vice versa, if a coronavirus adapts to human host and has low virulence, its HP score is high. For example, two high virulent CoVs, SARS-CoV (2003) \citep{marra2003genome} has low HP score 4.5139. The SARS-CoV related CoVs (SLCoV/RsSHC014, SLCoV/Rs3367, SLCoV/MA15, SLCoV/WIV1,SLCoV/Shaanxi2011, civets-SCoV/SZ3) have similar low HP scores as SARS-CoV. The highly lethal MERS-CoV (2012) \citep{van2012genomic} has very low HP score 3.2432. \textit{Rousettus} bat coronavirus bat-CoV/HKU9-1 (2011) \citep{lau2010coexistence}, closely related to MERS-CoV, has HP score 3.0114. Note that highly pathogenic Swine acute diarrhea syndrome coronavirus (SADS-CoV) (2017) \citep{zhou2018fatal} with potential human infection \citep{edwards2020swine} has low HP score 5.0192. The SAD-CoV is closely related and shares 95\% sequence identity to SLCoV/HKU2 (2004) \citep{zhou2018fatal}, which has low HP score 4.9346. Additionally, we measure the HP disparity of non-coronavirus Ebola virus, which is very virulent with a mortality rate of 51\% {kucharski2014case,di2020viral}. The HP score of Ebola virus is very low 2.5927. In brief, the low HP scores of the CoV genomes can be considered as the native state of the viruses in the original wild world. 

Since the 1960s when the first human coronavirus (HCoV) strain (HCoV-B814, 1965) was identified from a patient’s nasal discharge, over 30 different HCoV strains have been isolated, the most notable of which include HCoV-229E, HCoV-NL63, HCoV-HKU1, and HCoV-0C43 \citep{ye2020zoonotic}. Human CoVs (HCoV-229E, HCoV-OC43, HCoV-NL63, and HCoV-HKU1) are found worldwide and cause common cold predominantly in winter and spring \citep{poutanen2018human}. The HP computational result shows that Human-CoVs (229E (1963), NL63 (2004), and HKU-1 (2004)) have high HP scores (6.6788, 9.4172, and 10.4088, respectively). Note that HCoV-OC43 (1963) has low HP score 5.7846. The fact that HCoV-OC43 is actually virulent CoV to human may account for the low HP score. The colds due to HCoV-OC43 are more severe and are indistinguishable from colds due to rhinoviruses \citep{poutanen2018human}. The clinical symptoms of HCoV-OC43 outbreak in France include severe vomiting, diarrhea, abdominal pain, pneumonia, and even infection in the brain \citep{vabret2003outbreak}. Previously studies underscored the virulence of human-CoV/OC43 in elderly populations \citep{patrick2006outbreak} and Human-CoV/OC43 has serological cross-reactivity with SARS-CoV. Molecular clocks of Human-CoV/OC43 speculated that Human-CoV/OC43 caused the 1889–1890 flu pandemic\citep{vijgen2005complete}. Hence the low $R_{HP(AT/CG)}$ correlates the virulence of human-CoV/OC43. The high HP score of the CoV genome may be due to zoonotic adaptive infection and evolution, and indicates relative light virulence of the virus. 

From Dec. 2019 and ongoing, Severe acute respiratory syndrome coronavirus 2 (SARS-CoV-2, 2019) is a highly transmissible and pathogenic coronavirus of COVID-19, and the morbidity and mortality of the SARS-CoV-2 are less than SARS-CoV or MERS-CoV. The two coronaviruses SARS-CoV/Tor2 and MERS-CoV are very virulent for human host, and originated from bat host \citep{hemida2014mers}. Accordingly, SARS-CoV-2 has a higher HP score 6.0860 than SARS-CoV and MERS-CoV's. The three SARS-CoV-2 close related CoVs, SLCoV/RaTG13 \citep{zhou2020pneumonia}, SLCoV/BANAL-52 \citep{temmam2021coronaviruses} and bat-CoV/RmYN02 \citep{zhou2021identification} have a slightly large HP scores 6.2465, 6.3412, and 6.1050, respectively. Once Pangolin was considered as an intermediate animal host for SARS-CoV-2, Pangolin-SLCoV (2020) \citep{lam2020identifying} has HP score 6.8144, the value is much higher than the HP score of SARS-CoV-2. 

The zoonotic evolution of a coronavirus is the period of its infecting humans or closely related hosts from the wild state. In the wild state when the virus is with the native bat, we speculate that the HP score of the virus genome can be stable and relatively small. When the coronavirus infects human or mammalian hosts, under the host immune response and natural selection, the virus genome may undergo mutations, therefore resulting in an increased HP score for survival in the host. We speculate that the HP scores in CoV genomes are increasing during human host-interaction evolution. Additionally, \cite{denver2004abundance} previous study showed that the frequencies and run-length distributions of HP AT greatly outnumber HP GC in eukaryotic genome of \textit{Caenorhabditis elegans}. Yet the implication of the Homopolymeric nucleotides in virus genome evolution is little known. The HP distributions of the genomes of SARS-related CoVs in this study agree with the observations found in the eukaryotic genomes.

In summary, we observe the correlation between the HP disparity values $R_{HP(AT/CG)}$ and the virulence of coronaviruses. Higher $R_{HP(AT/CG)}$ values indicate lower virulence. This can be explained that when a bat coronavirus of wild status, because the virus is new to human host, then human immune system may have a strong reaction as cytokine storm on the infection of the wild virus, causing high pathogenicity \citep{mehta2020covid}. If the viruses already infected human for some periods, the virus genomes undergo multiple mutations, especially C>T mutations, as the result of the human APOBECs RNA editing. The resulting virus genomes will have more homopolymeric nucleotide T repeats, indicated as higher $R_{HP(AT/CG)}$. After some periods of virus infection and genome mutations, the virus adapts to the human immune system, and the human immune system reactions become less severe on the virus infection, therefore, the virulence of the virus is reduced. 

\begin{table}[ht]
	\caption{The HP scores of SARS-Like coronaviruses and Ebola virus.}
	\centering 
	\begin{tabular}{lllll}
		\hline\hline
		\noalign{\vskip 0.05in}   
		Virus & GenBank & $(A+T)/(C+G)$ & $R_{HP(AT/CG)}$ \\	
		\hline
		\noalign{\vskip 0.05in}   
		SARS-CoV-2 & NC\_045512& 1.6335 & 6.0860 \\ 
		SLCoV/RaTG13 & MN996532& 1.6295 & 6.2465 \\
		SLCoV/BANAL-52& EPI\_ISL\_4302644& 1.6338 & 6.3412 \\
		bat-CoV/RmYN02 & EPI\_ISL\_412977& 1.6151 & 6.1050\\
		Pangolin-SLCoV/GD/1& MT084071 & 1.6109& 6.8144\\
		SARS-CoV/BJ01 &AY278488&1.4513 & 4.5139\\ 
		SARS-CoV/Tor2 &AY274119&1.4533 & 4.5400\\
		MERS-CoV &NC\_019843& 1.4250 & 3.2432\\
		bat-CoV/HKU9-1 &EF065513& 1.4362 & 3.0114\\    
		SLCoV/ZXC21 &MG772934& 1.5758 & 5.3249\\ 
		SLCoV/ZC45 &MG772933& 1.5705 & 5.3898\\ 
		SLCoV/HKU2&NC\_009988& 1.5457& 4.9346\\
		SLCoV/RsSHC014 &KC881005 & 1.4445& 4.5484\\
		SLCoV/Rs3367&KC881006& 1.4512& 4.7344\\
		SLCoV/MA15&JF292916& 1.4495& 4.5422\\
		SLCoV/WIV1&KF367457& 1.4530& 4.7085\\
		SLCoV/Shaanxi2011 & JX993987& 1.4043& 4.0818\\
		civets-SCoV/SZ3 & AY304486& 1.4492& 4.5400\\
		swine-SADS-CoV & MG557844& 1.5403& 5.0192\\ 
		Human-CoV/229E & NC\_002645& 1.6136& 6.6788\\
		Human-CoV/NL63 &NC\_005831& 1.9018 & 9.4172\\
		Human-CoV/OC43 &AY585229& 1.7188 &5.7846\\
		Human-CoV/HKU1 &AY597011& 2.1192& 10.4088\\
		Ebola virus&KP120616& 1.4272& 2.5927\\
		\hline\hline
		\noalign{\vskip 0.05in}   
	\end{tabular}
	\label{table:nonlin} 	
\end{table}

\subsection{SARS-CoV-2 genomes have increasing HPs during zoonotic evolution}
The HP distributions in the genomes of bat and human CoVs suggest that the genome HPs of a CoV may indicate the trend of zoonotic evolution of the CoV. Therefore, we expect that the HP scores in SARS-CoV-2 isolates increase during the infection and evolution period from 2020 to 2021. To verify this expectation, we measure and compare the $R_{HP(AT/CG)}$ scores of SARS-CoV-2 isolates in global territories from different periods. The results show that the SARS-CoV-2 isolates in all test countries have increased HP scores over time. Fig.3 is the HP score trends of SARS-CoV-2 isolates in China, South Africa, and USA in 2020 and 2021 (as of 12/3/2021). Fig.3 indicates that SARS-CoV-2 isolates in China, South Africa, and USA (New York) keep increasing HPs during infection and evolution. For instance, the average HP score SARS-CoV-2 isolates in China and South Africa is 6.0937 in Jan./Feb. 2020, and is increased to 6.1586 in Nov./Dec., 2021 (Fig.3 (a) (b) and Table 2). A similar HP trend is shown in the SARS-CoV-2 isolates in USA (New York) (Fig.3(c)(d) and Table 3), the HP score is increased from 6.0868 in Mar./April, 2020 to 6.1455 in Nov./Dec., 2021. Additionally, similar HP score trends of SARS-CoV-2 isolates are found in India, Sweden, and Israel (in Supplementary Materials). 

\begin{figure}[tbp]
	\centering
	\subfloat[]{\includegraphics[width=2.75in]{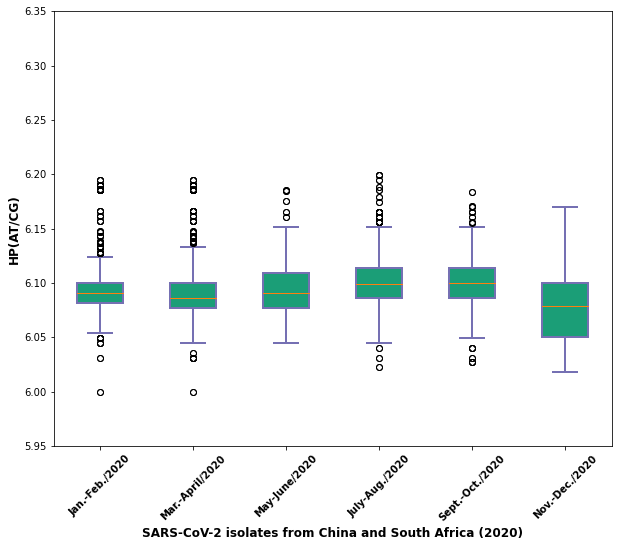}}
	\subfloat[]{\includegraphics[width=2.75in]{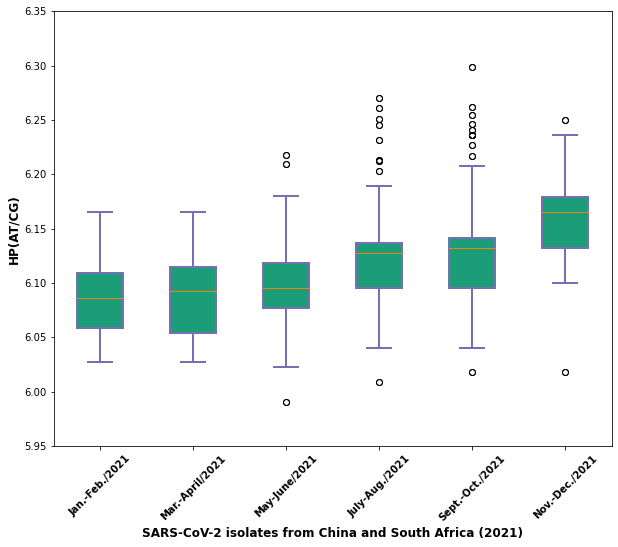}}\quad
	\subfloat[]{\includegraphics[width=2.75in]{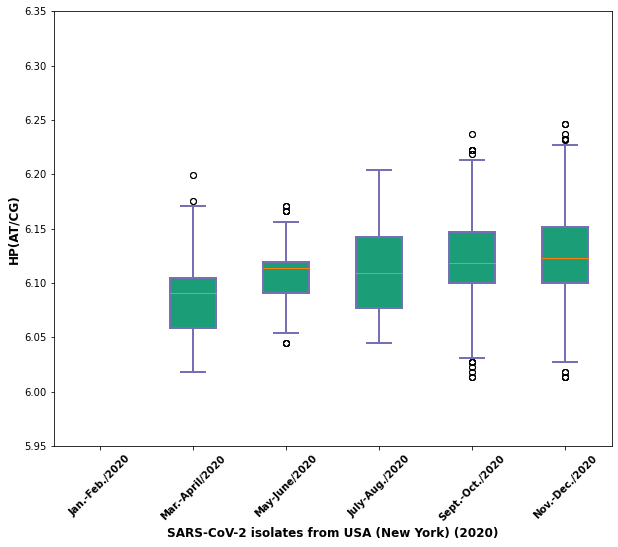}}
	\subfloat[]{\includegraphics[width=2.75in]{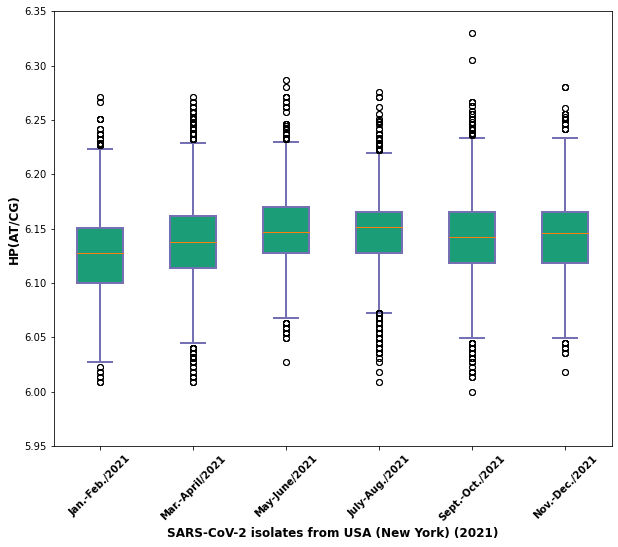}}\quad
	\caption{Boxplot of $R_{HP(AT/CG)}$ of SARS-CoV-2 isolates from different periods. (a) China and South Africa (2020), (b) China and South Africa (2021, as of Dec. 3,2021), (c) USA (New York, 2020), (d) USA (New York, 2021, as of Dec. 3, 2021).}
	\label{fig:sub1}
\end{figure}

\begin{table}[ht]
	\caption{$R_{HP(AT/CG)}$ scores of SARS-CoV-2 isolates in China and South Africa (2020-2021).}
	\centering 
	\begin{tabular}{llllll}
		\hline\hline
		\noalign{\vskip 0.05in}   
		period & sample size & max & min & mean & std \\	
		\hline
		\noalign{\vskip 0.05in} 
		Jan.-Feb./2020 &  587& 6.1944& 6.0000& 6.0937& 0.0239\\
		Mar.-April/2020 &822& 6.1944& 6.0000& 6.0909& 0.0239\\
		May-June/2020 & 134& 6.1852 &6.0450& 6.0939& 0.028\\
		July-Aug./2020&753& 6.1991& 6.0224& 6.0989& 0.0252\\
		Sept.-Oct./2020&336& 6.1835 &6.0269 &6.1007& 0.0256\\
		Nov.-Dec./2020&130& 6.1697& 6.0179& 6.0791& 0.0329\\
		Jan.-Feb./2021&33& 6.1651& 6.0269& 6.0885& 0.0328\\
		Mar.-April/2021&76& 6.1651& 6.0269& 6.0904& 0.0363\\
		May-June/2021&207& 6.2176 &5.9909& 6.0980& 0.0344\\
		July-Aug./2021&375& 6.2698 &6.0089& 6.1206& 0.0337\\
		Sept.-Oct./2021&340 &6.2991& 6.0179 &6.1269 &0.0411\\
		Nov.-Dec./2021&21& 6.2500& 6.0179& 6.1586& 0.0528\\
		\hline\hline
		\noalign{\vskip 0.05in}   
	\end{tabular}
	\label{table:nonlin} 	
\end{table}

\begin{table}[ht]
	\caption{$R_{HP(AT/CG)}$ scores of SARS-CoV-2 isolates in USA (New York) (2020-2021).}
	\centering 
	\begin{tabular}{llllll}
		\hline\hline
		\noalign{\vskip 0.05in}   
		period & sample size & max & min & mean & std \\	
		\hline
		\noalign{\vskip 0.05in}   
		Mar.-April/2020 & 2378 & 6.1991 & 6.0179 & 6.0868 & 0.0271\\
		May-June/2020 & 181 & 6.1705 & 6.0450 & 6.1098 & 0.0256\\
		July-Aug./2020&264& 6.2037 &6.0450 &6.1092& 0.0392\\
		Sept.-Oct./2020&1541& 6.2372 &6.0135& 6.1216& 0.0347\\
		Nov.-Dec./2020&2567& 6.2465& 6.0135& 6.1248& 0.0344\\
		Jan.-Feb./2021&6988 & 6.2710& 6.0089& 6.1267 &0.0353\\
		Mar.-April/2021&16682 & 6.2710& 6.0089& 6.1381 &0.0356\\
		May-June/2021&2337 & 6.2864& 6.0269 &6.1473& 0.0332\\
		July-Aug./2021&8022 & 6.2757& 6.0089& 6.1455& 0.0318\\
		Sept.-Oct./2021&9647 & 6.3301& 6.0000 &6.1412& 0.0354\\
		Nov.-Dec./2021&2227 &6.2804 & 6.0179& 6.1455& 0.0349\\ 
		\hline\hline
		\noalign{\vskip 0.05in}   
	\end{tabular}
	\label{table:nonlin} 	
\end{table}

During infection and zoonotic evolution, SARS-CoV-2 can mutate to some Variants of Concern (VOC). On November 24, 2021, the Omicron variant of SARS-CoV-2 (B.1.1.529) was detected in mid of November 2021, South Africa \citep{pulliam2021increased}, and will spread more easily than the original SARS-CoV-2 virus. The Omicron variant and the initial SARS-CoV-2 isolates in Jan.-Feb/2020 are compared for HP distributions (Fig. 4(a)). The result shows that the HPs in Omicron rise much higher than the initial SARS-CoV-2 isolates. The mean HPs of Omicron variants is 6.2498, and the initial SARS-CoV-2 is 6.0860. The Omicron variant has higher HP than SLCoV/RaTG13 (Table 1). For example, the Omicron strain (${\rm{EPI\_ISL\_7509826|2021-11-25}}$, England) has HP as 6.3019. We hypothesis that the Omicron variant with high HPs may not be very virulent. Furthermore, we measure the HPs in the genomes of SARS-CoV and MERS-CoV strains that were collected in various sources during 2003-2004, and 2012-2013, respectively (Fig. 4(b)) (The GenBank access numbers of the SARS-CoVs and MERS-CoVs, and the HP statistics are in Supplementary Materials). The results further confirm that HP disparity scores are increased in SARS-CoVs and MERS-CoVs during evolution over time. 

\begin{figure}[tbp]
	\centering
	\subfloat[]{\includegraphics[width=2.75in]{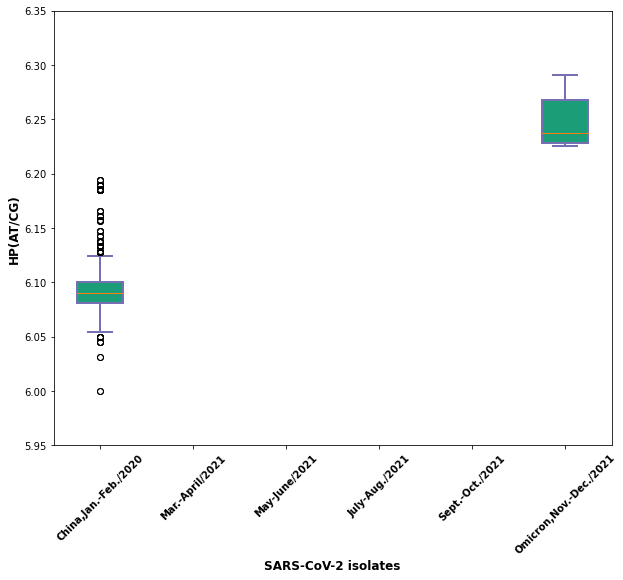}}
	\subfloat[]{\includegraphics[width=2.75in]{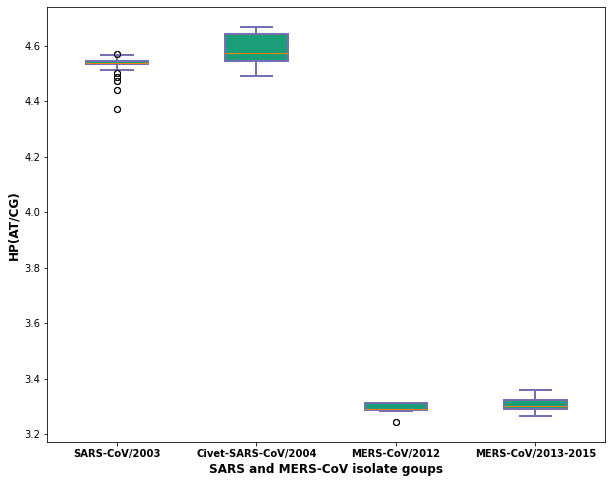}}\quad
	\caption{Boxplot of $R_{HP(AT/CG)}$ of SARS-CoV-2, SARS-CoV and MERS-CoV isolates in the globe from different periods. (a) SARS-CoV-2 (Jan.-Feb.,2020 and Omicron variants), (b) SARS-CoV and MERS-CoV.}
	\label{fig:sub1}
\end{figure}

To further analyze the correlation of polynucleotide compositions with the evolution of coronaviruses in human hosts, we plot the HPs (4-mer) along the genomes of SARS-like coronaviruses (SARS-CoVs) and typical beta-coronavirus human-CoV/OC43 (Fig.5). The results show that poly-T in human-CoV/OC43 is the highest along the CoV genomes (Fig.5(a)). This result is due to the fact human-CoV/43 has been infecting human hosts for long periods, while SARS-related coronaviruses are relatively new to human hosts. When infecting intermediate animals and human hosts, the host APOBEC immune system mutates C>T in the virus genome as a defense mechanism, therefore, more poly-T can occur and accumulate in the virus genomes over time. Among the tested SARS-CoVs, SARS-CoV-2 has the most poly-T tracts, indicating SARS-CoV-2 undergoes the highest C>T mutations. It is noted that three SARS-related coronaviruses SARS-CoV-2, RaTG13, and RmYN02 have almost the same poly-T tracts from position 1 to 11000 (the end position of nsp5), but different poly-T from 11000 to the end. In addition, RaTG13 and RmYN02 have very similar poly-T along the genomes. The poly-A pattern is similar to poly-T. The three coronaviruses, SARS-CoV-2, RaTG13, and RmYN02 have high poly-A tracts (Fig.5(b)). These results suggest that SARS-CoV-2's direct ancestry is RmYN02, followed by RaTG13 (Fig.5(a)). The longer a coronavirus stays in and adapts to human hosts, the higher poly-T may occur in the virus, and therefore lower virulence, and vice versa. For the poly-C and ploy-G tracts, MERS-Cov has the highest poly-C and poly-G contents (Fig.5(c) and Fig.5(d)). Because MERS-CoV is the severe virulent coronavirus, the high ploy-C is an indicator for the native status of the virus because few C>T mutations occurred in MERS-CoV during a very short zoonotic period. Hence, we may postulate that MERS-CoV is the youngest zoonotic coronavirus among SARS-CoVs. Moreover, Fig. 5(d) shows that more poly-Gs are in the S-protein gene regions, indicating the hot-spots of immune response action, or recombinant events in the S protein genes. 

Because ploy-T may be the consequence of C>T mutations of the APOBEC Cytidine-to-Uracil deamination during interacting with human hosts, the result suggests that SARS-CoV-2 has a longer evolution period in human hosts than SARS-CoV/Tor2. Equally, SARS-CoV-2 has been lived long in human hosts than SARS-CoV/Tor2, i.e., SARS-CoV-2 is older in the zoonotic age than SARS-CoV/Tor2. Therefore, we may consider HP disparity of poly-(A/T) and poly-(C/G) as a measure for zoonotic age of the virus since its zoonotic infection. 

In summary, we see that the results from SARS-CoV-2 isolates, SARS-CoV and MERS-CoV are consistent with the observations that coronaviruses have been increasing $R_{HP(AT/CG)}$ values during zoonotic evolution time. These results can be possibly explained as follows. During SARS-CoV-2 infection, human immune system fights SARS-CoV-2 virus and more C>T mutations then generate homopolymeric nucleotide repeat Ts. Consequently, we may use the HP distributions to measure the zoonotic age or evolution trend of coronaviruses.

\begin{figure}[tbp]
	\centering
	\subfloat[]{\includegraphics[width=2.5in]{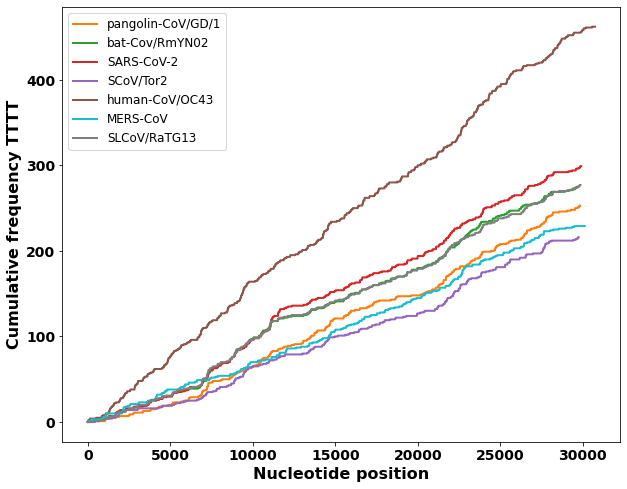}}
	\subfloat[]{\includegraphics[width=2.5in]{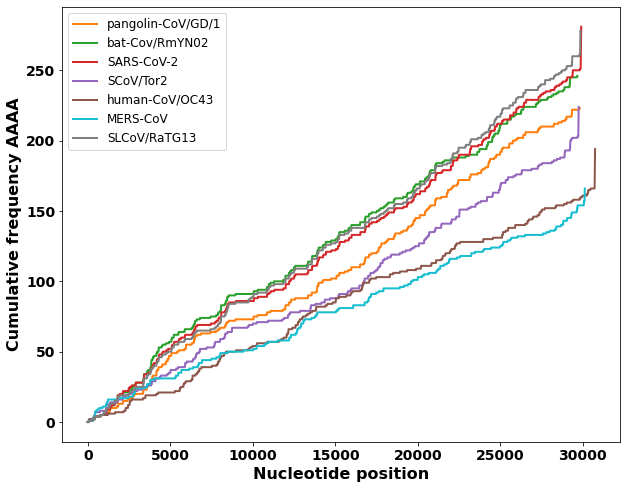}}\quad
	\subfloat[]{\includegraphics[width=2.5in]{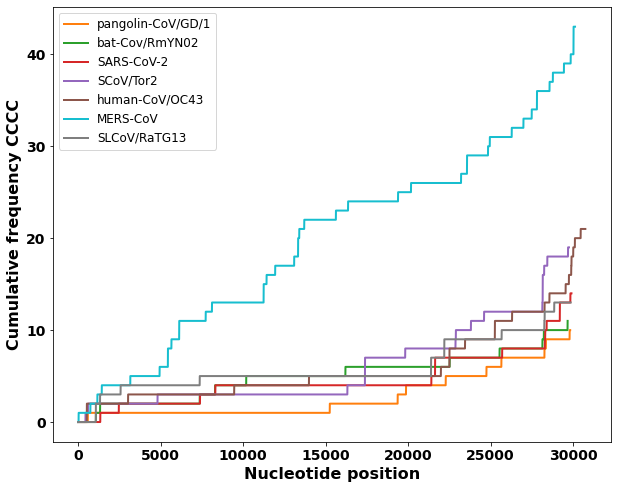}}
	\subfloat[]{\includegraphics[width=2.5in]{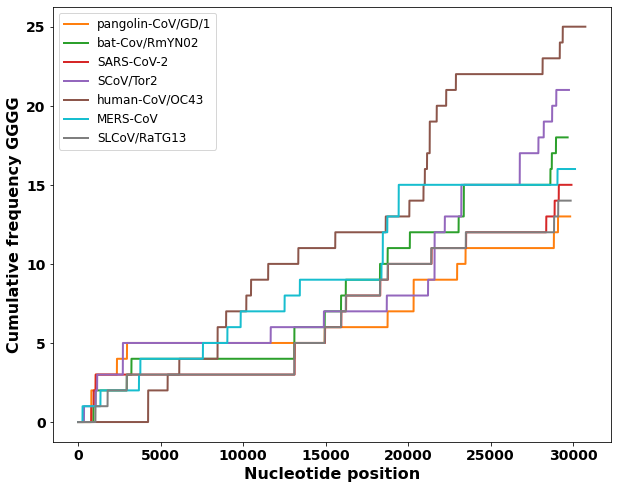}}\quad
	\caption{The cumulative frequencies of HPs along the SARS-CoV genomes. (a) The cumulative frequency of HP-T4. (b) The cumulative frequency of HP-A4. (c) The cumulative frequency of HP-C4. (d) The cumulative frequency of HP-G4.}
	\label{fig:sub1}
\end{figure}

\section{Discussions}
The COVID-19 pandemic has caused substantial health emergencies and economic stress in the world. Since there is no treatment efficacy for curing COVID-19, it is urgent to identify the zoonotic transmission of the causative agent SARS-CoV-2 so we may control and prevent future SARS-CoV-2 outbreaks. It is established that SARS-CoV, MERS-CoV, and SARS-CoV-2, cross the species barrier and move from bats to infect humans. Yet the necessity and mechanism that SARS-CoV-2 jumping from the intermediate animals to humans are unknown. One may argue that the bat coronaviruses can direct infect humans without being through the intermediate animals. If that is the case, the polynucleotide repeat analysis suggests that SARS-CoV-2 may have a similar or the same level of polynucleotide repeats as wild bat coronavirus. In fact, our results from show that SARS-CoV-2 has a similar level polynucleotide repeat level as bat-CoV/RmYN02, suggesting that SARS-CoV-2 may originate from bats and jump to human host directly, or, having long periods of adapting to human host before COVID-19 outbreak in Dec. 2019. 

From Table 1, we show that the zoonotic ages are correlated to the virulence of viruses because in theory when a virus fits more host immune system, under natural selection, the virulence becomes reduced \citep{alizon2009virulence}. Here, we postulate that the CoVs in wild bat host have a low $R_{HP(AT/CG)}$ value, and then more virulent to human host. After zoonotic infection to the human hosts, the virus genomes will have more homopolymeric nucleotide T repeats due to human immune system APOBECs RNA editing. We suggest that $R_{HP(AT/CG)}$ can be used as an approximate measure for the zoonotic age of the CoVs in human hosts. The values of $R_{HP(AT/CG)}$ from small to large correspond to the zoonotic evolution age from short to long. From the pAT/CG in Table 1 and C>T mutations as shown in Fig.2(c), we may presumably infer the order of the relative zoonotic ages of the coronaviruses from bat host to humans is as follows: MERS-CoV-->SARS-CoV/Tor2-->SARS-CoV-2-->bat-CoV/RmYN02-->SLCoV/RaTg13-->Pangolin-SLCoV. Here we see that SARS-CoV-2 is younger in human hosts than Pangolin-SLCoV. Therefore, we may infer the virulence of Pangolin-SLCoV shall be less than SARS-CoV-2. In addition, Table 1 shows bat coronavirus swine-SADS-CoV has low $R_{HP(AT/CG)}$ about 5. Note that this swine-SADS-CoV virus had caused millions of pig deaths in 2017, and low HP disparity score suggests the virus is in a native bat host state and much virulent to a new mammalian host.

When a human host is infected by a virus, two major deamination enzymes in human, ADAR and APOBEC, are responsible for adenosine-to-inosine deamination and cytidine-to-uracil deamination, leading to an observed A>G and C>T changes in the virus genomes, respectively. The consequence of Simple sequence repeats (SSRs) or microsatellites are DNA stretches consisting of short, tandem repeats of tri-, tetra- or penta-nucleotide HP smotifs. HPs are very special SSRs, and are ubiquitously dispersed in genomes. The SSR polymorphisms in bacterial pathogens are resulted from interactions with host structures and can be considered as a molecular marker of the genome evolution while infecting hosts \citep{moxon2006bacterial}.

The generation mechanism of the homopolymeric repeat G in SARS-CoV-2 genome is still unclear. One possible reason is that RNA-specific adenosine deaminase 1 (ADAR1) RNA editing adenosine-to-inosine deamination (A>G) may produce HP poly-G tracts \citep{picardi2022detection}. It is well established that sequences with G-blocks (adjacent runs of Guanines) can potentially form non-canonical G-quadruplex (G4) structures \citep{choi2011conformational,metifiot2014g}. For example, HP poly-G repeat is homopolymeric tract of Guanines (Gs) occurring in \textit{wafN} gene of \textit{Campylobacter jejuni} \citep{linton2000phase}, and 23 tandem repeats of the tetramer 5'-CAAT-3' present in \textit{lic2A} gene of \textit{Haemophilus influenzae}. The G4 structures are formed by stacking two or more G-tetrads by Hoogsteen hydrogen bonds and often are the sites of genomic instability, serving one or more biological functions \citep{bochman2012dna,chen2012sequence}. G-quadruplexes are also found in SARS-CoV genome. The non-structural protein 3 (nsp3) is one component of the viral replicase complex and contains a domain referred to as SARS unique domain (SUD), which interacts with G4s \citep{tan2009sars,kusov2015g}. The G4 nucleotides have been suggested in SARS-CoV-2 genomes \citep{frick2020molecular,ji2020discovery,zhang2020whole}. Therefore, poly-T and poly-G resulted from C>T and A>G mutations by immune response APOBEC and ADAR system, can be an evolution imprint that record the interaction between coronavirus and host. Here we propose to use the distribution of polynucleotide repeats for measuring the zoonotic evolution of coronaviruses \citep{hu2017discovery}. 

\section{Conclusions}
In this study, we analyze the SARS-CoV genomes and discover that SARS-CoV-2 genome is significantly enriched HP repeats. We propose the HP distributions of a CoV genome can be an effective indicator for the zoonotic evolution of CoV. We investigate the HP distributions in coronavirus genomes and present a method using the HP disparity as the measure of zoonotic evolution. The low HP disparity $R_{HP(AT/CG)}$ value of a coronavirus indicates that the virus is in native status in a bat and is far from infecting humans, therefore, the virus can be much virulent for humans. After a coronavirus infects humans, interacting with human immune system may produce more T bases, and also change the G4 structure landscape in the genome, therefore, increasing the HP disparity $R_{HP(AT/CG)}$ value.

SARS-CoV-2 emerged in Dec. 2019 as unexpected since SARS-CoV disappeared in 2004. From the outbreak of SARS-CoV-2, which is a recurrence of SARS-like CoV, it is expected that in the future more coronavirus outbreaks may occur, probably by different bat-CoV strains. Therefore, strict epidemic surveillance is indispensable for preventing future coronavirus outbreaks. The proposed polynucleotide repeat measure can be used in surveying the zoonotic evolution age and relative virulence of a given virus, therefore, it is of importance for tracking, virulence assessment, and controlling the outbreak of a coronavirus.

\section*{Abbreviations}
\begin{itemize}
	\item ADAR1: RNA-specific adenosine deaminase 1
	\item APOBEC: apolipoprotein B mRNA editing enzyme, catalytic polypeptide-like
	\item CoV: coronavirus
	\item COVID-19: coronavirus disease 2019 
	\item HP: homopolymeric nucleotide repeats
	\item MERS-CoV: middle east respiratory syndrome coronavirus
	\item SARS: severe acute respiratory syndrome
	\item SARS-CoV-2: severe acute respiratory syndrome coronavirus 2
	\item SADS-CoV: swine acute diarrhoea syndrome coronavirus
	\item SSR: simple sequence repeat
	\item WHO: the world health organization
\end{itemize}

\section*{Supplementary materials}
The supplementary materials are stored in the following public github repository.
https://github.com/cyinbox/GenomeSSR \\
\begin{itemize}
	\item GenBank complete sequences of the SARS related CoVs.
	\item HP disparity of SARS-CoV variants in Inida, Sweden, and Israel.
\end{itemize}
\clearpage
\bibliographystyle{elsarticle-harv}
\bibliography{myRefs}

\begin{thebibliography}{45}
\expandafter\ifx\csname natexlab\endcsname\relax\def\natexlab#1{#1}\fi
\providecommand{\url}[1]{\texttt{#1}}
\providecommand{\href}[2]{#2}
\providecommand{\path}[1]{#1}
\providecommand{\DOIprefix}{doi:}
\providecommand{\ArXivprefix}{arXiv:}
\providecommand{\URLprefix}{URL: }
\providecommand{\Pubmedprefix}{pmid:}
\providecommand{\doi}[1]{\href{http://dx.doi.org/#1}{\path{#1}}}
\providecommand{\Pubmed}[1]{\href{pmid:#1}{\path{#1}}}
\providecommand{\bibinfo}[2]{#2}
\ifx\xfnm\relax \def\xfnm[#1]{\unskip,\space#1}\fi
\bibitem[{Alizon et~al.(2009)Alizon, Hurford, Mideo and
  Van~Baalen}]{alizon2009virulence}
\bibinfo{author}{Alizon, S.}, \bibinfo{author}{Hurford, A.},
  \bibinfo{author}{Mideo, N.}, \bibinfo{author}{Van~Baalen, M.},
  \bibinfo{year}{2009}.
\newblock \bibinfo{title}{Virulence evolution and the trade-off hypothesis:
  history, current state of affairs and the future}.
\newblock \bibinfo{journal}{Journal of Evolutionary Biology}
  \bibinfo{volume}{22}, \bibinfo{pages}{245--259}.
\bibitem[{Bishop et~al.(2004)Bishop, Holmes, Sheehy and
  Malim}]{bishop2004apobec}
\bibinfo{author}{Bishop, K.N.}, \bibinfo{author}{Holmes, R.K.},
  \bibinfo{author}{Sheehy, A.M.}, \bibinfo{author}{Malim, M.H.},
  \bibinfo{year}{2004}.
\newblock \bibinfo{title}{Apobec-mediated editing of viral {RNA}}.
\newblock \bibinfo{journal}{Science} \bibinfo{volume}{305},
  \bibinfo{pages}{645--645}.
\bibitem[{Bochman et~al.(2012)Bochman, Paeschke and Zakian}]{bochman2012dna}
\bibinfo{author}{Bochman, M.L.}, \bibinfo{author}{Paeschke, K.},
  \bibinfo{author}{Zakian, V.A.}, \bibinfo{year}{2012}.
\newblock \bibinfo{title}{{DNA} secondary structures: stability and function of
  {G}-quadruplex structures}.
\newblock \bibinfo{journal}{Nature Reviews Genetics} \bibinfo{volume}{13},
  \bibinfo{pages}{770--780}.
\bibitem[{van Boheemen et~al.(2012)van Boheemen, de~Graaf, Lauber, Bestebroer,
  Raj, Zaki, Osterhaus, Haagmans, Gorbalenya, Snijder et~al.}]{van2012genomic}
\bibinfo{author}{van Boheemen, S.}, \bibinfo{author}{de~Graaf, M.},
  \bibinfo{author}{Lauber, C.}, \bibinfo{author}{Bestebroer, T.M.},
  \bibinfo{author}{Raj, V.S.}, \bibinfo{author}{Zaki, A.M.},
  \bibinfo{author}{Osterhaus, A.D.}, \bibinfo{author}{Haagmans, B.L.},
  \bibinfo{author}{Gorbalenya, A.E.}, \bibinfo{author}{Snijder, E.J.}, et~al.,
  \bibinfo{year}{2012}.
\newblock \bibinfo{title}{Genomic characterization of a newly discovered
  coronavirus associated with acute respiratory distress syndrome in humans}.
\newblock \bibinfo{journal}{MBio} \bibinfo{volume}{3},
  \bibinfo{pages}{e00473--12}.
\bibitem[{Chen(2020)}]{chen2020pathogenicity}
\bibinfo{author}{Chen, J.}, \bibinfo{year}{2020}.
\newblock \bibinfo{title}{Pathogenicity and transmissibility of {2019-nCoV} —
  a quick overview and comparison with other emerging viruses}.
\newblock \bibinfo{journal}{Microbes and Infection} \bibinfo{volume}{22},
  \bibinfo{pages}{69--71}.
\bibitem[{Chen and Yang(2012)}]{chen2012sequence}
\bibinfo{author}{Chen, Y.}, \bibinfo{author}{Yang, D.}, \bibinfo{year}{2012}.
\newblock \bibinfo{title}{Sequence, stability, structure of {G}-quadruplexes
  and their drug interactions}.
\newblock \bibinfo{journal}{Current Protocols in Nucleic Acid Chemistry/edited
  by Serge L. Beaucage [et al.]} , \bibinfo{pages}{Unit17--5}.
\bibitem[{Choi and Majima(2011)}]{choi2011conformational}
\bibinfo{author}{Choi, J.}, \bibinfo{author}{Majima, T.}, \bibinfo{year}{2011}.
\newblock \bibinfo{title}{Conformational changes of {non-B} {DNA}}.
\newblock \bibinfo{journal}{Chemical Society Reviews} \bibinfo{volume}{40},
  \bibinfo{pages}{5893--5909}.
\bibitem[{Denver et~al.(2004)Denver, Morris, Kewalramani, Harris, Chow, Estes,
  Lynch and Thomas}]{denver2004abundance}
\bibinfo{author}{Denver, D.R.}, \bibinfo{author}{Morris, K.},
  \bibinfo{author}{Kewalramani, A.}, \bibinfo{author}{Harris, K.E.},
  \bibinfo{author}{Chow, A.}, \bibinfo{author}{Estes, S.},
  \bibinfo{author}{Lynch, M.}, \bibinfo{author}{Thomas, W.K.},
  \bibinfo{year}{2004}.
\newblock \bibinfo{title}{Abundance, distribution, and mutation rates of
  homopolymeric nucleotide runs in the genome of {C}aenorhabditis elegans}.
\newblock \bibinfo{journal}{Journal of Molecular Evolution}
  \bibinfo{volume}{58}, \bibinfo{pages}{584--595}.
\bibitem[{Di~Giorgio et~al.(2020)Di~Giorgio, Martignano, Torcia, Mattiuz and
  Conticello}]{di2020evidence}
\bibinfo{author}{Di~Giorgio, S.}, \bibinfo{author}{Martignano, F.},
  \bibinfo{author}{Torcia, M.G.}, \bibinfo{author}{Mattiuz, G.},
  \bibinfo{author}{Conticello, S.G.}, \bibinfo{year}{2020}.
\newblock \bibinfo{title}{Evidence for host-dependent {RNA} editing in the
  transcriptome of {SARS-CoV-2}}.
\newblock \bibinfo{journal}{Science Advances} \bibinfo{volume}{6},
  \bibinfo{pages}{eabb5813}.
\bibitem[{Edwards et~al.(2020)Edwards, Yount, Graham, Leist, Hou, Dinnon, Sims,
  Swanstrom, Gully, Scobey et~al.}]{edwards2020swine}
\bibinfo{author}{Edwards, C.E.}, \bibinfo{author}{Yount, B.L.},
  \bibinfo{author}{Graham, R.L.}, \bibinfo{author}{Leist, S.R.},
  \bibinfo{author}{Hou, Y.J.}, \bibinfo{author}{Dinnon, K.H.},
  \bibinfo{author}{Sims, A.C.}, \bibinfo{author}{Swanstrom, J.},
  \bibinfo{author}{Gully, K.}, \bibinfo{author}{Scobey, T.D.}, et~al.,
  \bibinfo{year}{2020}.
\newblock \bibinfo{title}{Swine acute diarrhea syndrome coronavirus replication
  in primary human cells reveals potential susceptibility to infection}.
\newblock \bibinfo{journal}{Proceedings of the National Academy of Sciences}
  \bibinfo{volume}{117}, \bibinfo{pages}{26915--26925}.
\bibitem[{Frick et~al.(2020)Frick, Virdi, Vuksanovic, Dahal and
  Silvaggi}]{frick2020molecular}
\bibinfo{author}{Frick, D.N.}, \bibinfo{author}{Virdi, R.S.},
  \bibinfo{author}{Vuksanovic, N.}, \bibinfo{author}{Dahal, N.},
  \bibinfo{author}{Silvaggi, N.R.}, \bibinfo{year}{2020}.
\newblock \bibinfo{title}{Molecular basis for {ADP}-ribose binding to the
  {M}ac1 domain of {SARS-CoV-2 Nsp3}}.
\newblock \bibinfo{journal}{Biochemistry} .
\bibitem[{Hemida et~al.(2014)Hemida, Chu, Poon, Perera, Alhammadi, Ng, Siu,
  Guan, Alnaeem and Peiris}]{hemida2014mers}
\bibinfo{author}{Hemida, M.G.}, \bibinfo{author}{Chu, D.K.},
  \bibinfo{author}{Poon, L.L.}, \bibinfo{author}{Perera, R.A.},
  \bibinfo{author}{Alhammadi, M.A.}, \bibinfo{author}{Ng, H.y.},
  \bibinfo{author}{Siu, L.Y.}, \bibinfo{author}{Guan, Y.},
  \bibinfo{author}{Alnaeem, A.}, \bibinfo{author}{Peiris, M.},
  \bibinfo{year}{2014}.
\newblock \bibinfo{title}{{MERS} coronavirus in dromedary camel herd, {S}audi
  {A}rabia}.
\newblock \bibinfo{journal}{Emerging Infectious Diseases} \bibinfo{volume}{20},
  \bibinfo{pages}{1231}.
\bibitem[{Hu et~al.(2017)Hu, Zeng, Yang, Ge, Zhang, Li, Xie, Shen, Zhang, Wang
  et~al.}]{hu2017discovery}
\bibinfo{author}{Hu, B.}, \bibinfo{author}{Zeng, L.P.}, \bibinfo{author}{Yang,
  X.L.}, \bibinfo{author}{Ge, X.Y.}, \bibinfo{author}{Zhang, W.},
  \bibinfo{author}{Li, B.}, \bibinfo{author}{Xie, J.Z.}, \bibinfo{author}{Shen,
  X.R.}, \bibinfo{author}{Zhang, Y.Z.}, \bibinfo{author}{Wang, N.}, et~al.,
  \bibinfo{year}{2017}.
\newblock \bibinfo{title}{Discovery of a rich gene pool of bat {SARS}-related
  coronaviruses provides new insights into the origin of {SARS} coronavirus}.
\newblock \bibinfo{journal}{PLoS pathogens} \bibinfo{volume}{13},
  \bibinfo{pages}{e1006698}.
\bibitem[{Janulczyk et~al.(2010)Janulczyk, Masignani, Maione, Tettelin, Grandi
  and Telford}]{janulczyk2010simple}
\bibinfo{author}{Janulczyk, R.}, \bibinfo{author}{Masignani, V.},
  \bibinfo{author}{Maione, D.}, \bibinfo{author}{Tettelin, H.},
  \bibinfo{author}{Grandi, G.}, \bibinfo{author}{Telford, J.L.},
  \bibinfo{year}{2010}.
\newblock \bibinfo{title}{Simple sequence repeats and genome plasticity in
  {S}treptococcus agalactiae}.
\newblock \bibinfo{journal}{Journal of Bacteriology} \bibinfo{volume}{192},
  \bibinfo{pages}{3990--4000}.
\bibitem[{Ji et~al.(2020)Ji, Juhas, Tsang, Kwok, Li and
  Zhang}]{ji2020discovery}
\bibinfo{author}{Ji, D.}, \bibinfo{author}{Juhas, M.}, \bibinfo{author}{Tsang,
  C.M.}, \bibinfo{author}{Kwok, C.K.}, \bibinfo{author}{Li, Y.},
  \bibinfo{author}{Zhang, Y.}, \bibinfo{year}{2020}.
\newblock \bibinfo{title}{Discovery of {G}-quadruplex-forming sequences in
  {SARS-CoV-2}}.
\newblock \bibinfo{journal}{Briefings in Bioinformatics} .
\bibitem[{Kusov et~al.(2015)Kusov, Tan, Alvarez, Enjuanes and
  Hilgenfeld}]{kusov2015g}
\bibinfo{author}{Kusov, Y.}, \bibinfo{author}{Tan, J.},
  \bibinfo{author}{Alvarez, E.}, \bibinfo{author}{Enjuanes, L.},
  \bibinfo{author}{Hilgenfeld, R.}, \bibinfo{year}{2015}.
\newblock \bibinfo{title}{A {G}-quadruplex-binding macrodomain within the
  “{SARS}-unique domain” is essential for the activity of the
  {SARS}-coronavirus replication--transcription complex}.
\newblock \bibinfo{journal}{Virology} \bibinfo{volume}{484},
  \bibinfo{pages}{313--322}.
\bibitem[{Lam et~al.(2020)Lam, Jia, Zhang, Shum, Jiang, Zhu, Tong, Shi, Ni,
  Liao et~al.}]{lam2020identifying}
\bibinfo{author}{Lam, T.T.Y.}, \bibinfo{author}{Jia, N.},
  \bibinfo{author}{Zhang, Y.W.}, \bibinfo{author}{Shum, M.H.H.},
  \bibinfo{author}{Jiang, J.F.}, \bibinfo{author}{Zhu, H.C.},
  \bibinfo{author}{Tong, Y.G.}, \bibinfo{author}{Shi, Y.X.},
  \bibinfo{author}{Ni, X.B.}, \bibinfo{author}{Liao, Y.S.}, et~al.,
  \bibinfo{year}{2020}.
\newblock \bibinfo{title}{Identifying {SARS-CoV-2-related} coronaviruses in
  {Malayan} pangolins}.
\newblock \bibinfo{journal}{Nature} \bibinfo{volume}{583},
  \bibinfo{pages}{282--285}.
\bibitem[{Lau et~al.(2010)Lau, Poon, Wong, Wang, Huang, Xu, Guo, Li, Gao, Chan
  et~al.}]{lau2010coexistence}
\bibinfo{author}{Lau, S.K.}, \bibinfo{author}{Poon, R.W.},
  \bibinfo{author}{Wong, B.H.}, \bibinfo{author}{Wang, M.},
  \bibinfo{author}{Huang, Y.}, \bibinfo{author}{Xu, H.}, \bibinfo{author}{Guo,
  R.}, \bibinfo{author}{Li, K.S.}, \bibinfo{author}{Gao, K.},
  \bibinfo{author}{Chan, K.H.}, et~al., \bibinfo{year}{2010}.
\newblock \bibinfo{title}{Coexistence of different genotypes in the same bat
  and serological characterization of {R}ousettus bat coronavirus {HKU9}
  belonging to a novel betacoronavirus subgroup}.
\newblock \bibinfo{journal}{Journal of Virology} \bibinfo{volume}{84},
  \bibinfo{pages}{11385--11394}.
\bibitem[{Linton et~al.(2000)Linton, Gilbert, Hitchen, Dell, Morris, Wakarchuk,
  Gregson and Wren}]{linton2000phase}
\bibinfo{author}{Linton, D.}, \bibinfo{author}{Gilbert, M.},
  \bibinfo{author}{Hitchen, P.G.}, \bibinfo{author}{Dell, A.},
  \bibinfo{author}{Morris, H.R.}, \bibinfo{author}{Wakarchuk, W.W.},
  \bibinfo{author}{Gregson, N.A.}, \bibinfo{author}{Wren, B.W.},
  \bibinfo{year}{2000}.
\newblock \bibinfo{title}{Phase variation of a $\beta$-1, 3
  galactosyltransferase involved in generation of the ganglioside {GM1}-like
  lipo-oligosaccharide of {C}ampylobacter jejuni}.
\newblock \bibinfo{journal}{Molecular Microbiology} \bibinfo{volume}{37},
  \bibinfo{pages}{501--514}.
\bibitem[{Marra et~al.(2003)Marra, Jones, Astell, Holt, Brooks-Wilson,
  Butterfield, Khattra, Asano, Barber, Chan et~al.}]{marra2003genome}
\bibinfo{author}{Marra, M.A.}, \bibinfo{author}{Jones, S.J.},
  \bibinfo{author}{Astell, C.R.}, \bibinfo{author}{Holt, R.A.},
  \bibinfo{author}{Brooks-Wilson, A.}, \bibinfo{author}{Butterfield, Y.S.},
  \bibinfo{author}{Khattra, J.}, \bibinfo{author}{Asano, J.K.},
  \bibinfo{author}{Barber, S.A.}, \bibinfo{author}{Chan, S.Y.}, et~al.,
  \bibinfo{year}{2003}.
\newblock \bibinfo{title}{The genome sequence of the {SARS}-associated
  coronavirus}.
\newblock \bibinfo{journal}{Science} \bibinfo{volume}{300},
  \bibinfo{pages}{1399--1404}.
\bibitem[{Maty{\'a}{\v{s}}ek and
  Kova{\v{r}}{\'\i}k(2020)}]{matyavsek2020mutation}
\bibinfo{author}{Maty{\'a}{\v{s}}ek, R.}, \bibinfo{author}{Kova{\v{r}}{\'\i}k,
  A.}, \bibinfo{year}{2020}.
\newblock \bibinfo{title}{Mutation patterns of human {SARS-CoV-2} and bat
  {RaTG13} coronavirus genomes are strongly biased towards {C>U} transitions,
  indicating rapid evolution in their hosts}.
\newblock \bibinfo{journal}{Genes} \bibinfo{volume}{11}, \bibinfo{pages}{761}.
\bibitem[{Mehta et~al.(2020)Mehta, McAuley, Brown, Sanchez, Tattersall, Manson,
  Collaboration et~al.}]{mehta2020covid}
\bibinfo{author}{Mehta, P.}, \bibinfo{author}{McAuley, D.F.},
  \bibinfo{author}{Brown, M.}, \bibinfo{author}{Sanchez, E.},
  \bibinfo{author}{Tattersall, R.S.}, \bibinfo{author}{Manson, J.J.},
  \bibinfo{author}{Collaboration, H.A.S.}, et~al., \bibinfo{year}{2020}.
\newblock \bibinfo{title}{{COVID-19}: consider cytokine storm syndromes and
  immunosuppression}.
\newblock \bibinfo{journal}{Lancet (London, England)} \bibinfo{volume}{395},
  \bibinfo{pages}{1033}.
\bibitem[{M{\'e}tifiot et~al.(2014)M{\'e}tifiot, Amrane, Litvak and
  Andreola}]{metifiot2014g}
\bibinfo{author}{M{\'e}tifiot, M.}, \bibinfo{author}{Amrane, S.},
  \bibinfo{author}{Litvak, S.}, \bibinfo{author}{Andreola, M.L.},
  \bibinfo{year}{2014}.
\newblock \bibinfo{title}{{G}-quadruplexes in viruses: function and potential
  therapeutic applications}.
\newblock \bibinfo{journal}{Nucleic Acids Research} \bibinfo{volume}{42},
  \bibinfo{pages}{12352--12366}.
\bibitem[{Monchatre-Leroy et~al.(2017)Monchatre-Leroy, Bou{\'e}, Boucher,
  Renault, Moutou, Ar~Gouilh and Umhang}]{monchatre2017identification}
\bibinfo{author}{Monchatre-Leroy, E.}, \bibinfo{author}{Bou{\'e}, F.},
  \bibinfo{author}{Boucher, J.M.}, \bibinfo{author}{Renault, C.},
  \bibinfo{author}{Moutou, F.}, \bibinfo{author}{Ar~Gouilh, M.},
  \bibinfo{author}{Umhang, G.}, \bibinfo{year}{2017}.
\newblock \bibinfo{title}{Identification of alpha and beta coronavirus in
  wildlife species in france: bats, rodents, rabbits, and hedgehogs}.
\newblock \bibinfo{journal}{Viruses} \bibinfo{volume}{9}, \bibinfo{pages}{364}.
\bibitem[{Moxon et~al.(2006)Moxon, Bayliss and Hood}]{moxon2006bacterial}
\bibinfo{author}{Moxon, R.}, \bibinfo{author}{Bayliss, C.},
  \bibinfo{author}{Hood, D.}, \bibinfo{year}{2006}.
\newblock \bibinfo{title}{Bacterial contingency loci: the role of simple
  sequence {DNA} repeats in bacterial adaptation}.
\newblock \bibinfo{journal}{Annu. Rev. Genet.} \bibinfo{volume}{40},
  \bibinfo{pages}{307--333}.
\bibitem[{Nelson et~al.(2003)Nelson, Carnegie, Martin, Ejtehadi, Hooley, Roden,
  Rowland-Jones, Warren, Astley and Murray}]{nelson2003demystified}
\bibinfo{author}{Nelson, P.N.}, \bibinfo{author}{Carnegie, P.},
  \bibinfo{author}{Martin, J.}, \bibinfo{author}{Ejtehadi, H.D.},
  \bibinfo{author}{Hooley, P.}, \bibinfo{author}{Roden, D.},
  \bibinfo{author}{Rowland-Jones, S.}, \bibinfo{author}{Warren, P.},
  \bibinfo{author}{Astley, J.}, \bibinfo{author}{Murray, P.G.},
  \bibinfo{year}{2003}.
\newblock \bibinfo{title}{Demystified... human endogenous retroviruses}.
\newblock \bibinfo{journal}{Molecular Pathology} \bibinfo{volume}{56},
  \bibinfo{pages}{11}.
\bibitem[{Patrick et~al.(2006)Patrick, Petric, Skowronski, Guasparini, Booth,
  Krajden, McGeer, Bastien, Gustafson, Dubord et~al.}]{patrick2006outbreak}
\bibinfo{author}{Patrick, D.M.}, \bibinfo{author}{Petric, M.},
  \bibinfo{author}{Skowronski, D.M.}, \bibinfo{author}{Guasparini, R.},
  \bibinfo{author}{Booth, T.F.}, \bibinfo{author}{Krajden, M.},
  \bibinfo{author}{McGeer, P.}, \bibinfo{author}{Bastien, N.},
  \bibinfo{author}{Gustafson, L.}, \bibinfo{author}{Dubord, J.}, et~al.,
  \bibinfo{year}{2006}.
\newblock \bibinfo{title}{An outbreak of human coronavirus {OC43} infection and
  serological cross-reactivity with {SARS} coronavirus}.
\newblock \bibinfo{journal}{Canadian Journal of Infectious Diseases and Medical
  Microbiology} \bibinfo{volume}{17}.
\bibitem[{Picardi et~al.(2022)Picardi, Mansi and Pesole}]{picardi2022detection}
\bibinfo{author}{Picardi, E.}, \bibinfo{author}{Mansi, L.},
  \bibinfo{author}{Pesole, G.}, \bibinfo{year}{2022}.
\newblock \bibinfo{title}{Detection of { A-to-I} {RNA} editing in
  {SARS-COV-2}}.
\newblock \bibinfo{journal}{Genes} \bibinfo{volume}{13}, \bibinfo{pages}{41}.
\bibitem[{Poutanen(2018)}]{poutanen2018human}
\bibinfo{author}{Poutanen, S.M.}, \bibinfo{year}{2018}.
\newblock \bibinfo{title}{Human coronaviruses}.
\newblock \bibinfo{journal}{Principles and Practice of Pediatric Infectious
  Diseases} , \bibinfo{pages}{1148}.
\bibitem[{Pulliam et~al.(2021)Pulliam, van Schalkwyk, Govender, von Gottberg,
  Cohen, Groome, Dushoff, Mlisana and Moultrie}]{pulliam2021increased}
\bibinfo{author}{Pulliam, J.R.}, \bibinfo{author}{van Schalkwyk, C.},
  \bibinfo{author}{Govender, N.}, \bibinfo{author}{von Gottberg, A.},
  \bibinfo{author}{Cohen, C.}, \bibinfo{author}{Groome, M.J.},
  \bibinfo{author}{Dushoff, J.}, \bibinfo{author}{Mlisana, K.},
  \bibinfo{author}{Moultrie, H.}, \bibinfo{year}{2021}.
\newblock \bibinfo{title}{Increased risk of {SARS-CoV-2} reinfection associated
  with emergence of the omicron variant in {S}outh {A}frica}.
\newblock \bibinfo{journal}{medRxiv} .
\bibitem[{Shu and McCauley(2017)}]{shu2017gisaid}
\bibinfo{author}{Shu, Y.}, \bibinfo{author}{McCauley, J.},
  \bibinfo{year}{2017}.
\newblock \bibinfo{title}{{GISAID}: {G}lobal initiative on sharing all
  influenza data--from vision to reality}.
\newblock \bibinfo{journal}{Eurosurveillance} \bibinfo{volume}{22}.
\bibitem[{Tan et~al.(2009)Tan, Vonrhein, Smart, Bricogne, Bollati, Kusov,
  Hansen, Mesters, Schmidt and Hilgenfeld}]{tan2009sars}
\bibinfo{author}{Tan, J.}, \bibinfo{author}{Vonrhein, C.},
  \bibinfo{author}{Smart, O.S.}, \bibinfo{author}{Bricogne, G.},
  \bibinfo{author}{Bollati, M.}, \bibinfo{author}{Kusov, Y.},
  \bibinfo{author}{Hansen, G.}, \bibinfo{author}{Mesters, J.R.},
  \bibinfo{author}{Schmidt, C.L.}, \bibinfo{author}{Hilgenfeld, R.},
  \bibinfo{year}{2009}.
\newblock \bibinfo{title}{The {SARS}-unique domain ({SUD}) of {SARS}
  coronavirus contains two macrodomains that bind {G}-quadruplexes}.
\newblock \bibinfo{journal}{PLoS pathogens} \bibinfo{volume}{5}.
\bibitem[{Temmam et~al.(2021)Temmam, Vongphayloth, Salazar, Munier, Bonomi,
  R{\'e}gnault, Douangboubpha, Karami, Chretien, Sanamxay
  et~al.}]{temmam2021coronaviruses}
\bibinfo{author}{Temmam, S.}, \bibinfo{author}{Vongphayloth, K.},
  \bibinfo{author}{Salazar, E.B.}, \bibinfo{author}{Munier, S.},
  \bibinfo{author}{Bonomi, M.}, \bibinfo{author}{R{\'e}gnault, B.},
  \bibinfo{author}{Douangboubpha, B.}, \bibinfo{author}{Karami, Y.},
  \bibinfo{author}{Chretien, D.}, \bibinfo{author}{Sanamxay, D.}, et~al.,
  \bibinfo{year}{2021}.
\newblock \bibinfo{title}{Coronaviruses with a {SARS-CoV-2}-like
  receptor-binding domain allowing {ACE2}-mediated entry into human cells
  isolated from bats of {I}ndochinese peninsula} .
\bibitem[{Vabret et~al.(2003)Vabret, Mourez, Gouarin, Petitjean and
  Freymuth}]{vabret2003outbreak}
\bibinfo{author}{Vabret, A.}, \bibinfo{author}{Mourez, T.},
  \bibinfo{author}{Gouarin, S.}, \bibinfo{author}{Petitjean, J.},
  \bibinfo{author}{Freymuth, F.}, \bibinfo{year}{2003}.
\newblock \bibinfo{title}{An outbreak of coronavirus {OC43} respiratory
  infection in {Normandy, France}}.
\newblock \bibinfo{journal}{Clinical Infectious Diseases} \bibinfo{volume}{36},
  \bibinfo{pages}{985--989}.
\bibitem[{Vijgen et~al.(2005)Vijgen, Keyaerts, Mo{\"e}s, Thoelen, Wollants,
  Lemey, Vandamme and Van~Ranst}]{vijgen2005complete}
\bibinfo{author}{Vijgen, L.}, \bibinfo{author}{Keyaerts, E.},
  \bibinfo{author}{Mo{\"e}s, E.}, \bibinfo{author}{Thoelen, I.},
  \bibinfo{author}{Wollants, E.}, \bibinfo{author}{Lemey, P.},
  \bibinfo{author}{Vandamme, A.M.}, \bibinfo{author}{Van~Ranst, M.},
  \bibinfo{year}{2005}.
\newblock \bibinfo{title}{Complete genomic sequence of human coronavirus
  {OC43}: molecular clock analysis suggests a relatively recent zoonotic
  coronavirus transmission event}.
\newblock \bibinfo{journal}{Journal of Virology} \bibinfo{volume}{79},
  \bibinfo{pages}{1595--1604}.
\bibitem[{Wang et~al.(2020a)Wang, Horby, Hayden and Gao}]{wang2020novel}
\bibinfo{author}{Wang, C.}, \bibinfo{author}{Horby, P.W.},
  \bibinfo{author}{Hayden, F.G.}, \bibinfo{author}{Gao, G.F.},
  \bibinfo{year}{2020}a.
\newblock \bibinfo{title}{A novel coronavirus outbreak of global health
  concern}.
\newblock \bibinfo{journal}{The Lancet} \bibinfo{volume}{395},
  \bibinfo{pages}{470--473}.
\bibitem[{Wang et~al.(2020b)Wang, Hozumi, Zheng, Yin and Wei}]{wang2020host}
\bibinfo{author}{Wang, R.}, \bibinfo{author}{Hozumi, Y.},
  \bibinfo{author}{Zheng, Y.H.}, \bibinfo{author}{Yin, C.},
  \bibinfo{author}{Wei, G.W.}, \bibinfo{year}{2020}b.
\newblock \bibinfo{title}{Host immune response driving {SARS-CoV-2} evolution}.
\newblock \bibinfo{journal}{Viruses} \bibinfo{volume}{12},
  \bibinfo{pages}{1095}.
\bibitem[{WHO(2021)}]{who_2021}
\bibinfo{author}{WHO}, \bibinfo{year}{2021}.
\newblock \bibinfo{title}{Who {C}oronavirus ({COVID-19}) dashboard}.
\newblock \bibinfo{journal}{Coronavirus Disease ({COVID-2019}) Situation
  Reports} \bibinfo{volume}{00}, \bibinfo{pages}{00--00}.
\bibitem[{Wu et~al.(2020)Wu, Zhao, Yu, Chen, Wang, Song, Hu, Tao, Tian, Pei
  et~al.}]{wu2020new}
\bibinfo{author}{Wu, F.}, \bibinfo{author}{Zhao, S.}, \bibinfo{author}{Yu, B.},
  \bibinfo{author}{Chen, Y.M.}, \bibinfo{author}{Wang, W.},
  \bibinfo{author}{Song, Z.G.}, \bibinfo{author}{Hu, Y.}, \bibinfo{author}{Tao,
  Z.W.}, \bibinfo{author}{Tian, J.H.}, \bibinfo{author}{Pei, Y.Y.}, et~al.,
  \bibinfo{year}{2020}.
\newblock \bibinfo{title}{A new coronavirus associated with human respiratory
  disease in {China}}.
\newblock \bibinfo{journal}{Nature} \bibinfo{volume}{579},
  \bibinfo{pages}{265--269}.
\bibitem[{Ye et~al.(2020)Ye, Yuan, Yuen, Fung, Chan and Jin}]{ye2020zoonotic}
\bibinfo{author}{Ye, Z.W.}, \bibinfo{author}{Yuan, S.}, \bibinfo{author}{Yuen,
  K.S.}, \bibinfo{author}{Fung, S.Y.}, \bibinfo{author}{Chan, C.P.},
  \bibinfo{author}{Jin, D.Y.}, \bibinfo{year}{2020}.
\newblock \bibinfo{title}{Zoonotic origins of human coronaviruses}.
\newblock \bibinfo{journal}{International Journal of Biological Sciences}
  \bibinfo{volume}{16}, \bibinfo{pages}{1686}.
\bibitem[{Zhang et~al.(2020)Zhang, Xiao, Gu, Liu and Sun}]{zhang2020whole}
\bibinfo{author}{Zhang, R.}, \bibinfo{author}{Xiao, K.}, \bibinfo{author}{Gu,
  Y.}, \bibinfo{author}{Liu, H.}, \bibinfo{author}{Sun, X.},
  \bibinfo{year}{2020}.
\newblock \bibinfo{title}{Whole genome identification of potential
  {G-quadruplexes} and analysis of the {G-quadruplex} binding domain for
  {SARS-CoV-2}}.
\newblock \bibinfo{journal}{bioRxiv} .
\bibitem[{Zhou et~al.(2020a)Zhou, Chen, Hu, Li, Song, Liu, Wang, Liu, Yang,
  Holmes et~al.}]{zhou2020novel}
\bibinfo{author}{Zhou, H.}, \bibinfo{author}{Chen, X.}, \bibinfo{author}{Hu,
  T.}, \bibinfo{author}{Li, J.}, \bibinfo{author}{Song, H.},
  \bibinfo{author}{Liu, Y.}, \bibinfo{author}{Wang, P.}, \bibinfo{author}{Liu,
  D.}, \bibinfo{author}{Yang, J.}, \bibinfo{author}{Holmes, E.C.}, et~al.,
  \bibinfo{year}{2020}a.
\newblock \bibinfo{title}{A novel bat coronavirus closely related to
  {SARS-CoV-2} contains natural insertions at the {S1/S2} cleavage site of the
  spike protein}.
\newblock \bibinfo{journal}{Current Biology} .
\bibitem[{Zhou et~al.(2021)Zhou, Ji, Chen, Bi, Li, Wang, Hu, Song, Zhao, Chen
  et~al.}]{zhou2021identification}
\bibinfo{author}{Zhou, H.}, \bibinfo{author}{Ji, J.}, \bibinfo{author}{Chen,
  X.}, \bibinfo{author}{Bi, Y.}, \bibinfo{author}{Li, J.},
  \bibinfo{author}{Wang, Q.}, \bibinfo{author}{Hu, T.}, \bibinfo{author}{Song,
  H.}, \bibinfo{author}{Zhao, R.}, \bibinfo{author}{Chen, Y.}, et~al.,
  \bibinfo{year}{2021}.
\newblock \bibinfo{title}{Identification of novel bat coronaviruses sheds light
  on the evolutionary origins of {SARS-CoV-2} and related viruses}.
\newblock \bibinfo{journal}{Cell} .
\bibitem[{Zhou et~al.(2018)Zhou, Fan, Lan, Yang, Shi, Zhang, Zhu, Zhang, Xie,
  Mani et~al.}]{zhou2018fatal}
\bibinfo{author}{Zhou, P.}, \bibinfo{author}{Fan, H.}, \bibinfo{author}{Lan,
  T.}, \bibinfo{author}{Yang, X.L.}, \bibinfo{author}{Shi, W.F.},
  \bibinfo{author}{Zhang, W.}, \bibinfo{author}{Zhu, Y.},
  \bibinfo{author}{Zhang, Y.W.}, \bibinfo{author}{Xie, Q.M.},
  \bibinfo{author}{Mani, S.}, et~al., \bibinfo{year}{2018}.
\newblock \bibinfo{title}{Fatal swine acute diarrhoea syndrome caused by an
  {HKU2}-related coronavirus of bat origin}.
\newblock \bibinfo{journal}{Nature} \bibinfo{volume}{556},
  \bibinfo{pages}{255--258}.
\bibitem[{Zhou et~al.(2020b)Zhou, Yang, Wang, Hu, Zhang, Zhang, Si, Zhu, Li,
  Huang et~al.}]{zhou2020pneumonia}
\bibinfo{author}{Zhou, P.}, \bibinfo{author}{Yang, X.L.},
  \bibinfo{author}{Wang, X.G.}, \bibinfo{author}{Hu, B.},
  \bibinfo{author}{Zhang, L.}, \bibinfo{author}{Zhang, W.},
  \bibinfo{author}{Si, H.R.}, \bibinfo{author}{Zhu, Y.}, \bibinfo{author}{Li,
  B.}, \bibinfo{author}{Huang, C.L.}, et~al., \bibinfo{year}{2020}b.
\newblock \bibinfo{title}{A pneumonia outbreak associated with a new
  coronavirus of probable bat origin}.
\newblock \bibinfo{journal}{Nature} \bibinfo{volume}{579},
  \bibinfo{pages}{270--273}.

\end{thebibliography}
\end{document}